\documentclass[12pt]{article} 
\usepackage{graphicx} 
\usepackage{amsmath,caption,threeparttable}  
\usepackage{newtxtext,newtxmath} 
\usepackage[T1]{fontenc} 
\usepackage{ae,aecompl}

\begin{document}
	
\title{{\bf On the evolution of a stellar system \\ 
		in the context of the virial equation}} 
\author{V.~Yu.~Terebizh\thanks{E-mail: valery@terebizh.ru} \\ 
	\small{Crimean Astrophysical Observatory}} 
	
\maketitle 
	
\begin{abstract}
The virial equation is used to clarify the nature of the dynamic evolution 
of a stellar system. Compared to the kinetic equation, it gives a deeper but 
incomplete description of the process of relaxation to a quasi-stationary 
state, which here means the fulfillment of the virial theorem. Analysis 
shows that the time to reach the virial equlibrium state $T_v$ is about 
two to three dozen dynamic time periods $T_d$. Namely, during $T_v$ the 
virial ratio, the mean harmonic radius, and the root-mean-square radius of 
the system fluctuate, and then the first two characteristics stabilize near 
their equilibrium values, while the root-mean-square radius continues to 
grow (possibly ad infinitum). This indicates a fundamentally different 
behavior of the moment of inertia of the system relative to the center of 
gravity and its potential energy, leading to the formation of a relatively 
small equilibrium core and an extended halo. 
\end{abstract} 
	

\section{Introduction} 

Theoretical considerations, reinforced in recent years by extensive numerical 
simulations, show that the dynamical evolution of a stellar system in its own 
gravitational field is characterized by three basic time scales.

The shortest of them, the {\it dynamic time} $T_d \sim(G\rho)^{-1/2}$, where 
$G$ is the gravitational constant and~$\rho$~-- the mass-average density of 
the system, is associated with large-scale motions of matter in the early 
stages of system evolution. This parameter is also called the {\it crossing 
time}, because the return time of a body that has fallen from the surface of 
a homogeneous ball of density $\rho$ into a hole passing along its diameter 
is equal to ${(3\pi/G\rho)}^{1/2}$. 

According to Jeans (1915, 1919), the subsequent relaxation of the system to 
a quasi-stationary state in the smoothed, so-called {\it regular} 
gravitational field is described by the {\it collisionless Boltzmann equation} 
for the distribution function $f({\bf r,v},t)$ in 6-dimensional phase space, 
$$
  \frac{\partial f}{\partial t} + {\bf v}\,\frac{\partial f}
  {\partial {\bf r}} - \frac{\partial \Phi}{\partial {\bf r}}\,
  \frac{\partial f}{\partial {\bf v}} = 0, 
  \eqno(1)
$$
supplemented by the Poisson equation for the conjoint potential 
$\Phi({\bf r},t)$ [Henon 1982; Binney \& Tremaine 2008]. The definition of 
quasi-stationary state is often associated with the {\it virial equation}, 
which is valid for a set of~$N$ gravitating points in the center of mass 
coordinate system: 
$$
  \frac{1}{2}\,\frac{d^2 J(t)}{dt^2} = 2K(t) + W(t), 
  \eqno(2)
$$
where 
$$ 
  \begin{array}{lll} 
  J(t)= \sum_1^N m_i {\bf r}_i^2 , &  \\ 
  & \\ 
  K(t) = \sum_1^N m_i {\bf v}_i^2/2, \quad  \mbox{and}  &  \\  
  & \\ 
  W(t) = -G \sum_{i=1}^{N-1} \sum_{j=i+1}^N \frac{m_i m_j}{|{\bf r}_i-
  {\bf r}_j|} & \\ 
  \end{array}    
  \eqno(3) 
$$
are, respectively, the moment of inertia of the system, its kinetic and 
potential energies, whereas $m_i$, ${\bf r}_i(t)$ and ${\bf v}_i(t)$ are 
the mass, radius-vector and the speed of the $i$-th star.\footnote{The 
virial equation as presented here follows from Eq.~(5.133) and the 
Lagrange-Jacobi identity (5.136) of Chandrasekhar~(1942) monograph.}  
The total mass $M = \sum m_i$ of the star system and its total 
energy $E$ are assumed to be given. If the motions occur in a limited 
region of space, then, averaging Eq.~(2) over time, we obtain the equality 
$2 \langle K \rangle + \langle W \rangle = 0$, called the {\it virial 
theorem} (Landau \& Lifshitz~1976). Stellar systems are not closed in space, 
but it is assumed that after some time a quasi-stationary state is reached, 
in which the left side of Eq.~(2) becomes negligible, so, marking the 
parameters in quasi-stationary state with asterisks, we can take 
$$
  2K_* + W_* = 0. 
  \eqno(4)
$$ 
Within the framework of the discussed here approach, one should understand 
the quasi-stationary state as the {\it virial equilibrium state} (VES). The 
characteristic time interval for reaching VES will be denoted as $T_v$. 
Some models of internal rearrangement of a system evolving from the virial 
to a true quasi-stationary equilibrium were studied by Levin, Pakter \& 
Rizzato~(2008) and Benetti et al.~ (2014). 

Finally, the third stage, the relaxation of the system's core towards the 
Maxwell-Boltzmann state, takes even more time~$T_r$. For half a century 
it was believed that this process is due solely to the {\it irregular} 
gravitational field of the system, which is defined as the difference 
between the real and smoothed fields (Ambartsumian~1938; Chandrasekhar 
1942; Spitzer 1987). Since the spatial density of stars in galaxies is low, 
the main contribution to the process is made by pair collisions (close 
passages) of stars; it is taken into account by the non-zero 
{\it collisional term} on the right side of Eq.~(1). An explicit 
representation of this term for systems with Coulomb or gravitational 
interaction was given by Landau (1937). Research in recent decades has 
associated relaxation to a more efficient process of {\it dynamic chaos} 
(Gurzadyan \& Savvidy 1984, 1986); the corresponding relaxation time 
$T_r \simeq N^{1/3} T_d$ (according to Rastorguev \& Sementsov 2006, 
the exponent is $1/5$). Thermodynamic equilibrium is never reached 
already due to the long-range nature of the gravitational force 
(Lynden-Bell~1967; Levin, Pakter \& Rizzato~2008; Levin et al.~2014; 
Benetti et al.~2014); in addition, the uncloseness of the system is 
manifested in its external parts. 

As regards the evolution at the second of the stages mentioned above, the 
nature of the observed fast "Maxwellization" of galaxies in a regular 
field remained unclear for a long time. The revival of research in this 
direction was initiated by Henon~(1964) and Lynden-Bell~(1967); the latter 
proposed an appropriate stochastic mechanism, as he called it, {\it violent 
relaxation}. In the current understanding, this implies the importance of 
collective processes in systems with long-range interaction (Shu 1978; 
Levin, Pakter~\& Rizzato 2008; Levin et~al. 2013; Gurzadyan~\& Kocharyan 
2009). On the other hand, numerical simulations, starting with van~Albada 
(1982) studies and up to Halle, Colombi \& Peirani (2019) and Sylos Labini 
\& Capuzzo-Dolcetta (2020) recent calculations, gradually clarify the 
commensurate role of radial instabilities and internal density fluctuations 
that lead to the formation of local substructures of increasing size. 
Unlike~$T_d$ and~$T_r$, no explicit representation of $T_v$ in terms of the 
integral parameters of the system has been found so far, especially since 
it depends on the initial state. Quantification is hindered by the extreme 
complexity of combining the kinetic and Poisson equations. 

In this connection, the virial equation attracts more attention. It 
should be taken to a deeper level of description compared to the kinetic 
equation in one its form or another, because the latter is inevitably 
formulated with finite accuracy, while the virial equation is due only 
to the fundamental fact that the potential energy in the gravitational 
interaction of a pair of point-like bodies is inversely proportional to 
the distance between them, i.e., it is a {\it homogeneous function} of 
coordinates of degree $-1$. Among other things, the virial equation is 
valid for any, small or large, number of interacting points, while the 
accuracy of Eq.~(1) drops as~$N$ decreases. For $N \gg 1$, the 
collisionless Boltzmann equation is consistent with the virial equation 
in the sense that Eq.~(2), in its continuous version, can be derived 
from Eq.~(1). 

Since the total energy of an isolated system $E = K(t)+W(t)$ is conserved 
in time, Eq.~(2) is usually written as 
$$
  \frac{1}{2}\,\frac{d^2 J(t)}{dt^2} = 2E - W(t).  
  \eqno(5)
$$
For a gravitationally bound system, a necessary (but not sufficient) 
stability condition is $E<0$ (Chandrasekhar 1942); we will assume that this 
condition is satisfied. Equation~(5) includes two unknown functions of time, 
and therefore, by itself, does not allow complete description of even the 
integral properties of the system. However, it can be hoped that it will 
provide some information about {\it the character} of evolution and the 
corresponding time intervals. The present paper is devoted to elucidating 
this possibility.

\section{Signs of different evolution of two characteristic radii 
	of the system} 

Usually the degree of proximity to the state of virial equilibrium 
is given by the value of the {\it virial ratio} 
$$
  V(t) \equiv \frac{2K(t)}{|W(t)|} = 2\left[ 1-\frac{E}{W(t)} \right], 
  \qquad 0 \le V < 2. 
  \eqno(6)
$$ 
In view of Eq.~(4), the values of the kinetic and potential energies in 
VES are  
$$
  K_* = -E,\qquad W_* = 2E, 
  \eqno(7)
$$
so the equilibrium virial ratio $V_* = 1$. As the time scale in VES, we 
take $T_*\equiv \mathfrak{R}_*/v_*$~-- the time of intersection of the 
mean harmonic radius $\mathfrak{R}_*$ of the system with the characteristic 
velocity $v_*$. The values of the last two follow from Eq.~(7) and the 
definitions 
$$
  K_* \equiv \frac{1}{2} Mv_*^2,\qquad   W_* \equiv 
  -\frac{GM^2}{2\mathfrak{R}_*}\,. 
  \eqno(8) 
$$ 
In this way we get: 
$$
  v_* = \left( \frac{2|E|}{M} \right)^{1/2}, \qquad 
  \mathfrak{R}_* = \frac{GM^2}{4|E|},\qquad 
  T_* = \frac{G}{4} \left( \frac{M^5}{2|E|^3} \right)^{1/2}. 
  \eqno(9)
$$ 
The literature uses similar definitions with minor differences in numerical 
coefficients; the values adopted here are convenient for what follows. 
Notice that all these parameters are determined by the values of $M$ and $E$. 
We can write the last of Eqs.~(9) as $T_* = (3/2\pi G \rho_*)^{1/2}$, where 
characteristic density $\rho_* \equiv 3M/4\pi \mathfrak{R}_*^3$; as expected, 
the time scale $T_*$ turns out to be of the order of dynamic time~$T_d$. 
Since the estimates of $v_*$ and $\mathfrak{R}_*$ can be found directly from 
observations, the first two of formulas~(9) make it possible to estimate the 
total mass and energy of the system (Bahcall~\& Tremaine 1981; Binney~\& 
Tremaine 2008). The discussed picture of evolution is illustrated in Fig.~1 
(see also Fig.~6.1 in Ciotti (2021) monograph). 

%% Figure 1 #############################################################
\begin{figure*} 
	\includegraphics[width=0.60\linewidth]{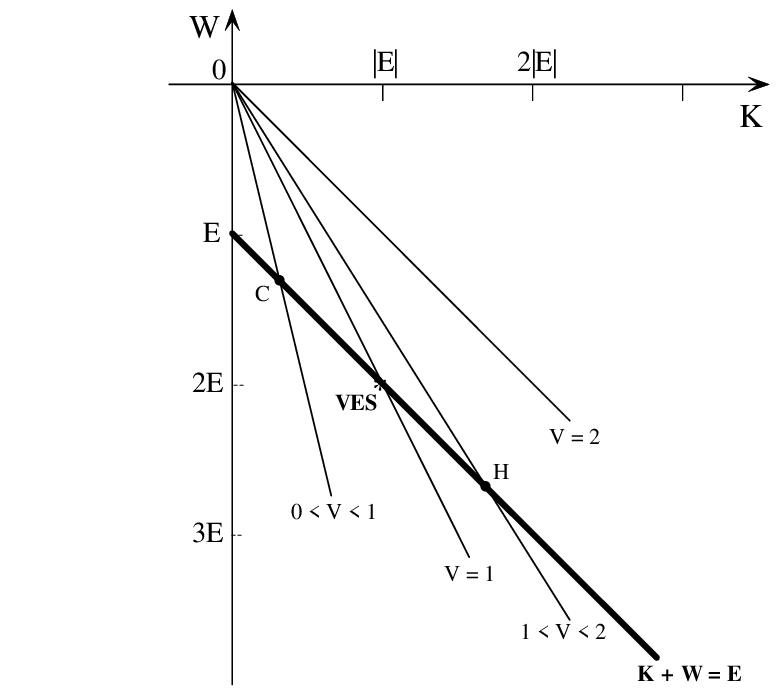}
	\caption{\small In the $(K,W)$ plane, the system evolves along the 
		straight line $K(t)+W(t)=E$, approaching on average a {\it virial 
			equilibrium state} (VES). The thin lines correspond to fixed 
		values of the virial ratio $V$. The letters $C$ and $H$ denote 
		``cold'' and ``hot'' configurations, which are characterized by 
		the values $K<|E|$ and $K>|E|$, respectively.
	}
	\label{fig01}  
\end{figure*}
%#######################################################################

It is convenient to pass in Eq.~(5) to the searched quantities of a single 
physical nature~-- the root-mean-square radius $R(t)$ and the mean harmonic 
radius $\mathfrak{R}(t)$, which are defined as follows: 
$$
  \begin{array}{ll} 
  R^2(t) \equiv \frac{1}{M} \sum_1^N m_i {\bf r}_i^2, & \\
  & \\
  \mathfrak{R}^{-1}(t) \equiv \frac{2}{M^2} \sum_{i=1}^{N-1} 
  \sum_{j=i+1}^N \frac{m_i m_j}{|{\bf r}_i-{\bf r}_j|} , & \\ 
  \end{array}
  \eqno(10)
$$
so that  
$$
  J(t) = M R^2(t),\qquad    W(t) = -\frac{GM^2}{2\,\mathfrak{R}(t)}\,,
  \eqno(11)
$$ 
and Eq.~(5) takes the form: 
$$
  \frac{d}{dt}\left(R\,\frac{dR}{dt}  \right) = 
  \frac{GM}{2\,\mathfrak{R}(t)} - \frac{2|E|}{M}\,.
  \eqno(12)
$$
Finally, introducing, with the help of Eq.~(9), the dimensionless variables 
$$
  \tau \equiv t/T_*, \qquad   x \equiv \mathfrak{R}/\mathfrak{R}_*,\qquad 
  \mbox{and}  \quad  y \equiv R/\mathfrak{R}_*,
  \eqno(13)
$$
we reduce the virial equation to the dimensionless form with all unit 
coefficients:
$$
  \frac{d}{d\tau}\left( y\,\frac{dy}{d\tau}  \right) = \frac{1}{x} - 1, 
  \eqno(14)
$$ 
while the definition~(6) for the virial ratio becomes 
$$
  V(\tau) = 2 - x(\tau),\qquad       0 < x(\tau) \le 2. 
  \eqno(15) 
$$

We emphasize an important fact that has not been discussed before: 
\begin{center}
\quad {\it The mean harmonic radius of a system with a negative total 
	energy \\ does not exceed twice the equilibrium value.} 
\end{center} 
The assertion follows from the definition $\mathfrak{R}/\mathfrak{R_*} 
\equiv 2|E|/|W|$ and the inequalities $K > 0$, $|E| < |W|$. A clear 
evidence is that the lines $V = \mbox{const}$ in Fig.~1 do not intersect 
with the line corresponding to the given value~$E$ when 
$\mbox{const} \ge 2$. 

%% Figure 2 ########################################################## 
\begin{figure*} 
	\includegraphics[width=0.60\linewidth]{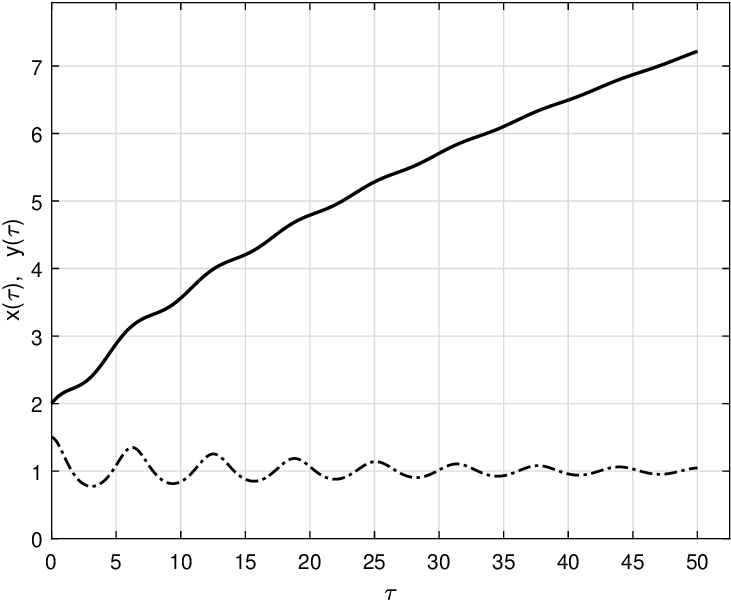} 
	\caption{\small 
	Change in the RMS radius of the system $y(\tau)$ (solid line) 
	for a given behavior of the harmonic radius $x(\tau)$ 
	(dash-dotted line). 
	} 
	\label{fig02} 
\end{figure*} 
%%####################################################################

In view of the above, the approximation to virial equilibrium over time 
means that the harmonic radius of the system $\mathfrak{R}(t)$, changing 
in a relatively narrow range of values $(0,2\mathfrak{R}_*)$, tends to 
its equilibrium value~(9), i.e., $x(\tau) \to 1$, the second derivative 
of $R^2(t)$ tends to zero, while the RMS radius $R(t)$ tends to a finite 
or infinite value. To show the theoretical possibility of the described 
scenario, we, anticipating the discussion in the next section, present in 
Fig.~2 the evolution of the harmonic and root-mean-square radii for a given 
behavior of the first radius, while the change in the second radius is 
calculated according to the virial equation (14). Specifically, the initial 
values $x(0) = 1.5$ and $y(0) = 2.0$ were set, and it was assumed that 
$x(\tau)$ tends to~$1$, experiencing an exponential decay and oscillations 
with a period $2\pi T_*$ (see Eq.~(23) below). Note that $y^2(\tau)$ grows 
linearly as $\tau \to \infty$.  

%%% Table 01 ######################################################
\begin{table*} 
	\caption{\small{Characteristics of systems with central symmetry 
	for various spatial density distributions $\rho(r)$. The following 
	designations are accepted: $\rho_0$~-- central density; $M$~-- 
	total mass; $\bar R$, $R$ and $\mathfrak{R}$~-- respectively, 
	mean, mean square and mean harmonic radii.} 
	}
	\label{tab:perf} 
	\small{ 
		\begin{tabular}{|c|c|c|c|c|c|c|} 
			\hline
			\rule[-1ex]{0pt}{4.0ex} {\bf  No.} & {\bf $\rho(r)/\rho_0$} & 
			{\bf $M/\rho_0  a^3$} & {\bf $\bar R/a$} & {\bf $R/a$} & 
			{\bf $\mathfrak{R}/a$} & {\bf $q \equiv R/\mathfrak{R}$} \\ 
			\hline 
			\rule[-1ex]{0pt}{4.0ex} 1 &1, $r \le a; \,\, 0, r>a$ & 
			$4\pi/3$ & $3/4$ & $\sqrt{3/5}$ & 5/6 & $2(3/5)^{3/2} 
			\simeq 0.9295$ \\ 
			\hline
			\rule[-1ex]{0pt}{4.0ex} 2 & $\exp(-r/a)$ & $8\pi $ & 3 & 
			$2\sqrt{3}$ & $16/5$ & $5\sqrt{3}/8 \simeq 1.0825$ \\ 
			\hline
			\rule[-1ex]{0pt}{4.0ex} 3 & $\exp(-r^2/a^2)$ & $\pi^{3/2}$ & 
			$2/\sqrt{\pi}$ & $\sqrt{3/2}$ & $\sqrt{\pi/2}$ & $\sqrt{3/\pi} 
			\simeq 0.9772 $\\
			\hline
			\rule[-1ex]{0pt}{4.0ex} 4 & $ (1+r^2/a^2)^{-5/2}$ & $4\pi/3$ & 
			$2$ & $\infty$& $16/3\pi$ & $\infty$ \\
			\hline 
			\rule[-1ex]{0pt}{4.0ex} 5 & $ (1+r^2/a^2)^{-2}$ & $\pi^2$ & 
			$\infty$ & $\infty$ & $\pi$ & $\infty$ \\ 
			\hline 
		\end{tabular} 
	}
\end{table*}  
%%%#############################################################

Supporting this model is the practical constancy of the half-mass 
radius $R_h$ in time when calculating the evolution of isolated stellar 
systems (Spitzer 1987). Within the framework of King's models, which 
describe well the internal structure of the globular clusters, the mean 
harmonic radius $\mathfrak{R} \simeq 2.5 R_h$ and, therefore, also 
changes little (private communication, A.~Rastorguev, 2023). 

Thus, the ratio of radii 
$$
q(\tau) \equiv    \frac{R}{\mathfrak{R}} = \frac{y(\tau)}{x(\tau)} 
\eqno(16)
$$
can vary within wide limits. It is useful to estimate the parameter $q$ 
for several continuous, for simplicity, density  distributions that differ 
significantly from each other. Table~1 lists the corresponding data for 
systems with central symmetry. These distributions can be conditionally 
considered as instantaneous (but not sequential) states of a star cluster 
evolving  in accordance with the kinetic and Poisson equations. To 
calculate the values given in the table, note that for a continuous 
distribution, the mass~$M(r)$ inside a sphere of radius~$r$ and the 
potential energy~$W$ are defined by the formulas 
$$ 
  \begin{array}{ll} 
  M(r) = 4\pi \int_0^r \rho(r) r^2 dr, & \\ 
  & \\ 
  W = -4\pi G \int_0^\infty \rho(r) M(r) r dr, & \\ 
  \end{array} 
  \eqno(17) 
$$
so the analogs of Eqs.~(10) are reduced, using Eq.~(11), to  
$$
  \begin{array}{ll} 
  R^2 = \frac{4\pi}{M} \int_0^\infty \rho(r) r^4 dr, & \\
  & \\
  \mathfrak{R}^{-1} = \frac{8\pi}{M^2} \int_0^\infty \rho(r) M(r)rdr. & \\ 
  \end{array} 
  \eqno(18)
$$ 

As Table~1 shows, for the first three, relatively homogeneous distributions, 
the values of $q$ are close to~$1$. Apparently, these density distributions 
are adequate only at the initial stage of evolution. The values of~$q$ 
remain of the same order for fairly significant deviations from the 
spherical symmetry of the density distribution, for example, towards 
ellipsoidality. More important is the density distribution in the outer 
region of the system. The theoretical estimates by von~Hoerner (1956), the 
thorough numerical modeling by van Albada (1982), reinforced by physical 
arguments of Trenti, Bertin \& van Albada (2005), further simulations of 
cluster evolution by Yangurazova \&~Bisnovatyi-Kogan (1984), Levin, Pakter 
\&~Rizzato (2008), Levin et al. (2013), Joyce, Marcos \&~Sylos 
Labini~(2010), Sylos Labini~(2013), Halle, Colombi \&~Peirani~(2019), and 
Sylos Labini \&~Capuzzo-Dolcetta~(2020) indicate the formation of a 
power-law density distribution $\rho(r) \propto r^{-\alpha}$ with exponent 
$\alpha \sim 3.3 - 4$ in the halo. The last two examples of Table~1 are 
just that. Example No.~4 is the system of Schuster~(1883) and 
Plummer~(1911), which was repeatedly used in connection with studies of 
globular star clusters. With density distributions as flat as in 
Examples~4 and~5, the integrals for~$R$ diverge, so the $q$-factor is 
infinitely large. 

The above models assume a more or less gradual change in density with 
distance from the center. An idea of the reverse behavior is given by a 
two-layer model with radii $R_1$, $R_2$ and densities $\rho_1$,  
$\rho_2$ in the central and outer zones, respectively. We do not present 
the corresponding formulas because of their cumbersomeness. The general 
conclusion is that at moderate values of the ratio $\rho_1/\rho_2$, the 
$q$-factor is still close to~1, and only when $\rho_1/\rho_2 \gg 1$ and 
$R_1/R_2 < 0.1$ can $q$ more than~$10$ be achieved. 

Thus, it seems likely that relatively homogeneous systems at the initial 
stage of evolution are characterized by~$q(\tau)$ of the order of~$1$, 
while the $q$-factor increases significantly in the course of further 
evolution as a dense core and extended halo of the system are formed.

\section{Solutions for a given ratio of radii}
 
In addition to the formal reason for the studying in Eq.~(12) the 
dimensionless ratio of two unknown radii $R(t)$ and $\mathfrak{R}(t)$, 
the function $q(\tau)$ plays an important role by setting the systematic 
behavior of the RMS radius averaged over fast oscillations with a period 
of the order of dynamic time~$T_d$ (see Appendix). 

In order to verify the oscillatory nature of the solutions of the virial 
equation, we write down Eq.~(14) as 
$$
  \frac{d}{d\tau}\left( y\,\frac{dy}{d\tau} \right) = 
  \frac{q(\tau)}{y} - 1,  
  \eqno(19)
$$
and assume first that the ratio of the radii does not change with time. 
Lynden-Bell~(1967) additionally linearized the corresponding equation; 
Chandrasekhar~\& Elbert (1972) found a rather complicated analytical 
solution to a non-linear equation. Written in parametric form, the exact 
solution of Eq.~(19) at $q(\tau) \equiv q_0$ describes a classical cycloid 
$$ 
  \left\{   \begin{array}{ll} 
  y = q_0 - a \cos\theta - b \sin\theta, & \\ 
  \tau = q_0 \theta - a \sin\theta - b (1-\cos\theta), &  
  0 \le \theta < \infty\,,
  \end{array}   \right. 
  \eqno(20) 
$$
whereas the mean harmonic radius of the system 
$$
  x(\tau) = y(\tau)/q_0 = 1 - (a/q_0) \cos\theta - (b/q_0) \sin\theta 
  \eqno(21) 
$$ 
oscillates around the equilibrium value $x_*=1$. The constants $a$ and $b$ 
are determined by the initial state: 
$$
  a = q_0 - y_0, \qquad  b = -y_0 \cdot y_0'.
  \eqno(22) 
$$
It is convenient to proceed from three initial values, namely, the pair 
$(y_0,y_0')$ and the virial ratio $V_0$; then, according to Eqs.~(15) 
and~(16), we have $x_0 = 2-V_0$ and $q_0 = y_0/x_0$. 

 %% Figure 3 ########################################################## 
 \begin{figure*}  
 	\includegraphics[width=0.55\linewidth]{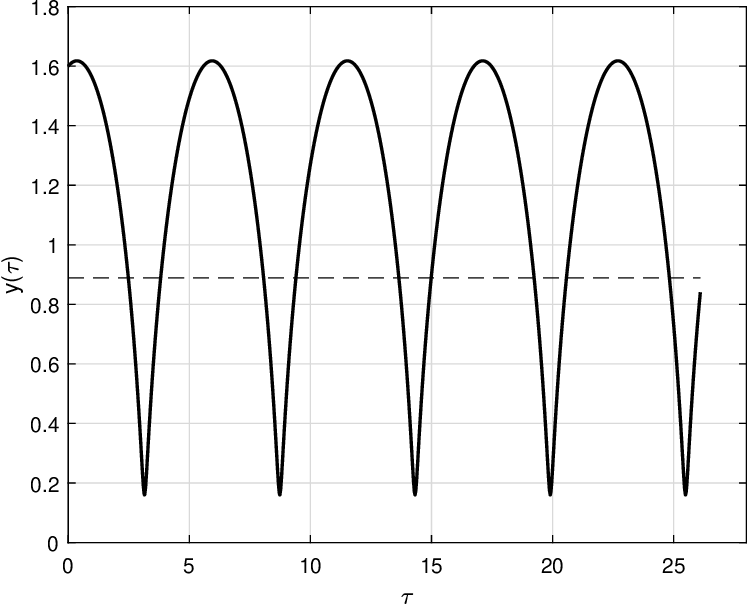} 
 	\caption{\small 
 	Changing the RMS radius $y(\tau)$ in a model with constant ratio 
 	$y/x \equiv q_0$ at initial values $y_0=1.60$, $y_0'=0.10$, 
 	$V_0=0.20$. The dashed line corresponds to $q_0=8/9$. 
 	} 
 	\label{fig03}  
 \end{figure*} 
 %%####################################################################
 
Equation~(6) shows that values of the kinetic energy less or greater than 
$|E|$ correspond to virial ratio values less or greater than~$1$ (see 
Fig.~1); how it is accepted, we call the respective states of the system 
``cold'' and ``hot''. The mean harmonic radius $\mathfrak{R}$ of the 
former state exceeds the equilibrium value $\mathfrak{R}_*$, while 
$\mathfrak{R} < \mathfrak{R}_*$  for the latter state. 
 
An example solution for $V_0=0.20$ (``cold'' system) is shown in Fig.~3. 
The period of oscillations of the cycloid in real time is equal to 
$q_0P_*$, where 
$$
  P_* = 2\pi T_* = \frac{\pi G}{2} \left( \frac{M^5}{2|E|^3} 
  \right)^{1/2} = \sqrt{6\pi/G\rho_*}, 
  \eqno(23)
$$  
and $\rho_* = 3M/4\pi \mathfrak{R}_*^3$ is the characteristic density 
of the cluster.

%% Figure 4 ##############################################################
\begin{figure*}  
	\begin{tabular}{cc}
		\includegraphics[height=4.5cm]{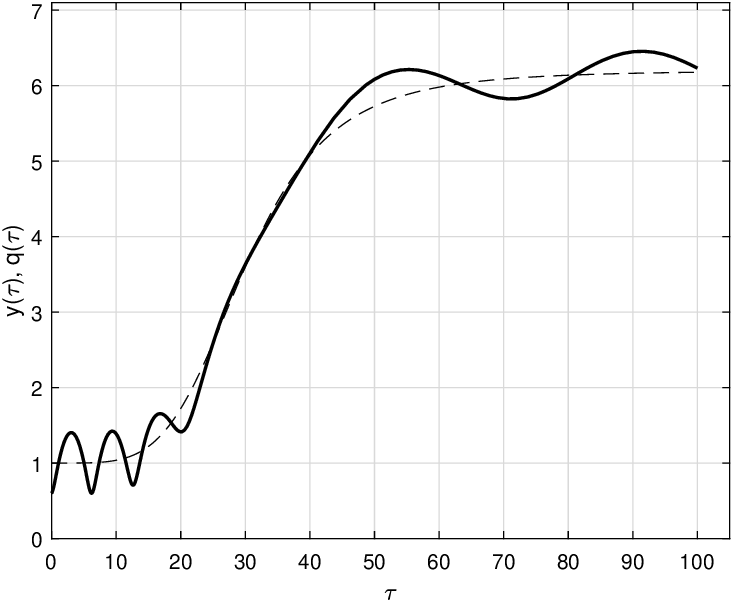} & 
		\includegraphics[height=4.5cm]{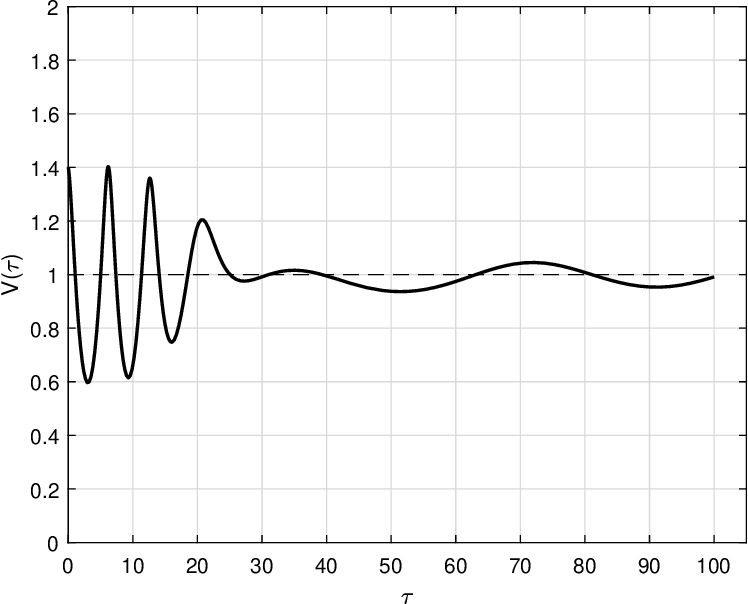} \\
	\end{tabular}  
	\caption{\small 
	Solution of Eq.~(19) with initial data $V_0 = 1.40$, $y_0=0.60$, 
	$y_0'=0.10$. Left: RMS radius $y(\tau)$ of the system (solid 
	line) and $q(\tau)$ function (dashed line). Right: Virial 
	ratio $V(\tau)$ (solid line) and unit level corresponding 
	to VES (dashed line). 
	} 
	\label{fig04}  
\end{figure*}  
%%######################################################################## 

It is clear that the undamped, so-called homologous oscillations give only  
a preliminary description of the early evolution of a self-gravitating 
system. As noted at the end of the previous section, to estimate the 
characteristic time to reach virial equilibrium, it is necessary to take 
into account a progressive macroscopic inhomogeneity of the system. 
Accordingly, we must turn to a model with a time-varying ratio of radii 
$R/\mathfrak{R}$. The Appendix to this paper shows that the approximate 
solution of Eq.~(19) with an arbitrary function $q(\tau)$, on which only 
the condition of its slow change on the time scale $T_*$ is imposed, is a 
generalization of the classical cycloid, namely: 
$$ 
\left\{   \begin{array}{ll} 
	y = u(\theta) - a(\theta) \cos\theta - b(\theta) \sin\theta, & \\ 
	\tau = \theta u(\theta) - a(\theta) \sin\theta - b(\theta) 
	(1-\cos\theta),  &  0 \le \theta < \infty. 
\end{array}   \right. 
\eqno(24) 
$$
Here, the base function $u(\theta)$ is given by the implicit equation 
$u = q(\theta\cdot u)$, and the variable coefficients $a(\theta)$ and 
$b(\theta)$ depend on $u(\theta )$, that is, they are also given by the 
function $q(\tau)$. In the case of $q(\tau) \equiv q_0$, we get $u = q_0$, 
the coefficients $a,b$ become constant, and we return to the model considered 
above. The physical meaning of representation (24) is that the functions 
$u(\theta),a(\theta)$ and $b(\theta)$ change, following $q(\tau)$, relatively 
slowly, while their trigonometric factors reflect precisely these rapid 
variations of the RMS radius around $q(\tau)$, and the mean harmonic radius 
and virial ratio~-- around the equilibrium value equal to~$1$. Our numerical 
examples show that in the immediate vicinity of the VES, the oscillation 
period slightly increases. 
 
%% Figure 5 ############################################################
\begin{figure*} 
	\begin{tabular}{cc}
 		\includegraphics[height=4.5cm]{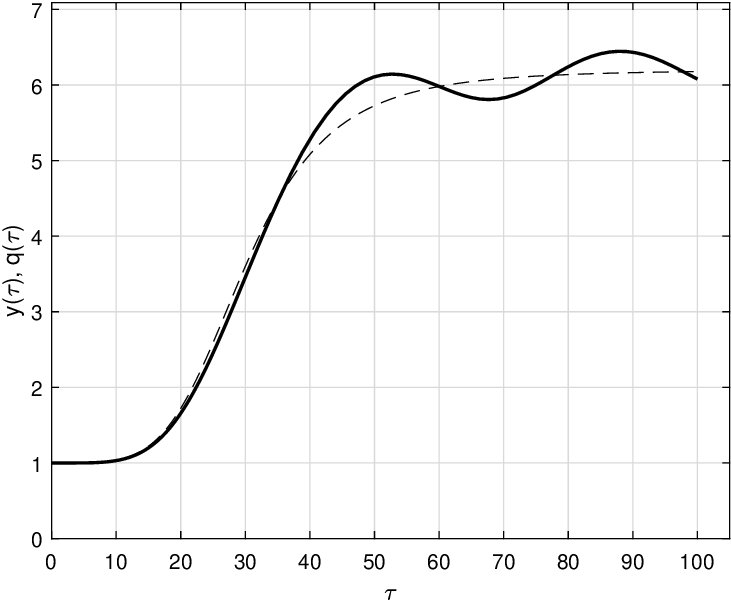} & 
 		\includegraphics[height=4.5cm]{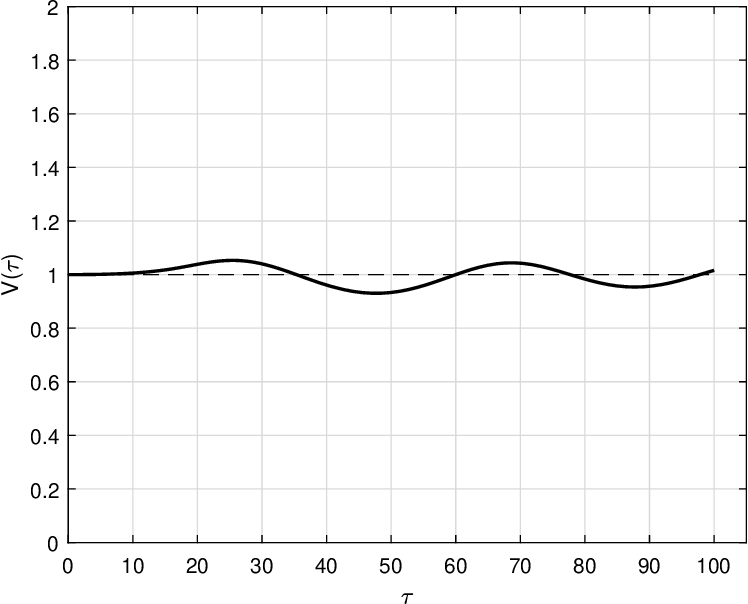} \\
 	\end{tabular} 
 	\caption{\small 
 	Solution of Eq.~(19) with initial data $V_0 = 1.0$, $y_0=1.0$, 
 	$y_0'=0$ corresponding to virial equilibrium state. Explanations 
 	are the same as in Fig.~4.
 	}  
 	\label{fig05} 
 \end{figure*}  
%% #####################################################################
 
Not as informative as the analytical, but more accurate approach is the 
direct numerical solution of Eq.~(19) for a given $q(\tau)$. Figure~4 gives 
an idea of a typical evolutionary pattern, in this case a ``hot'' system. 
The behavior of the mean harmonic radius $x(\tau)$ is not shown because, 
in view of Eq.~(15), it is a reflection of $V(\tau)$ relative to the 
equilibrium level. We see several rapid initial fluctuations in both the 
RMS radius and the virial ratio, however, later $y(\tau)$ follows the given 
function $q(\tau)$, while $V(\tau)$ and $x(\tau)$ practically stabilize 
around the equilibrium level $x_*=1$. {\it This means that, over a period 
of two to three dozen of dynamic time intervals $T_*$, an equilibrium core 
with radius $\mathfrak{R}_*$ is formed in the system, while the surrounding 
halo, which determines the RMS radius $R(t)$, continues to expand.} 

The pattern of virial oscillations seen in the right Fig.~4 has the same 
character as that obtained by numerical simulation of the dynamics of a 
multiparticle system (see, for example, Fig.~6 in Trenti, Bertin \& van 
Albada 2005). Fast oscillations are just a ringing against the backdrop 
of a slower reorganization of the system. Eventually, it is possible that 
after two dozen dynamic time periods $T_*$ the harmonic radius of the 
system $x(\tau)$ will become close to~$1$, and then, as the virial equation 
(14) shows, the further evolution of the root-mean-square radius of the 
system is described by simple law $R(t)/\mathfrak{R}_* \simeq q(\tau) 
\simeq (c_1\tau + c_2)^{1/ 2}$, where $\tau = t/T_*$ and $c_1 , c_2$ 
are some dimensionless constants.

For control, one should also check the evolution of the system, which was 
initially in a virial equilibrium state. As can be seen in Fig.~5, the 
rapid oscillations of the radii and the virial ratio have disappeared, 
the RMS radius $y(\tau)$ still follows $q(\tau)$, while the 
quasi-equilibrium nucleus experiences only long-term weak oscillations. 
This was to be expected. 

The difference in the behavior of the system core and halo seems quite 
plausible, but we should not forget that in the context considered here 
it is partly determined by the specification of the $q(\tau)$ function. 
This prompted us to consider models with different types of $q(\tau)$; 
all cases, except for extremely "hot" systems, show the same behavior.

\section{Concluding remarks} 
\label{sec:crem} 

The above analysis supports two features of the evolution of a stellar 
system towards virial equilibrium. First, its integral characteristics 
fluctuate during two to three tens of dynamic time~$T_*$ until the virial 
ratio stabilizes near the equilibrium value $V_*=1$. Secondly, the 
root-mean-square and mean harmonic radii vary in time in different ways, 
so the assumption previously accepted by a number of researchers about the 
approximate equality of radii is far from reality. As already noted, the 
first conclusion agrees with the results of numerical simulation of 
self-gravitating systems at $N \gg 1$. The second conclusion means a 
fundamentally different behavior of the moment of inertia of the system 
relative to the center of gravity and its potential energy. 

Further details of the process of approaching the virial equilibrium state 
remain hidden when only the virial equation is analyzed. Additional data 
are desirable, at least in the form of approximate relationships between 
the integral characteristics of the system. On the other hand, the 
difference between evolutionary paths of the RMS and the mean harmonic 
radii can be easily elucidated on the basis of both already performed and 
future numerical simulations.

\section*{Acknowledgements} 

The author is grateful to A.S. Rastorguev for useful comments.

\section*{Data availability} 

No new data were generated or analysed in support of this research.

\section*{Appendix. Approximate analytical solution of the virial equation} 

A nonlinear differential equation of the second order  
$$
  y \left[ \frac{d}{d\tau} \left( y\,\frac{dy}{d\tau}\right) + 1 \right] = 
  q(\tau)
  \eqno(A1)
$$
is considered in the domain $\tau \ge 0$ for a given non-negative function 
$q(\tau)$. The approach presented below, which goes back to the method of 
Van der Pol~(1927), is widely used in the theory of oscillations. 

We will look for a solution in a parametric form: 
$$ 
  \left\{   \begin{array}{ll} 
  y = u(\theta) - a(\theta) \cos\theta - b(\theta) \sin\theta, & \\ 
  \tau = \theta u(\theta) - a(\theta) \sin\theta - b(\theta) 
  (1-\cos\theta),  &  \theta \ge 0, \\ 
  \end{array}   \right. 
  \eqno(A2) 
$$ 
where $u(\theta)$, $a(\theta)$ and $b(\theta)$ are some unknown functions 
slowly varying over an interval of length $2\pi$. In particular, they can 
be constant. Specifically, we assume that $|u'(\theta)/u(\theta)| \ll 1$, 
and similar inequalities hold for the coefficients $a$ and $b$. Under this 
condition, the solution averaged over an interval of length $2\pi$ is 
$$
  \langle y(\theta) \rangle = \frac{1}{2\pi} 
  \int_{\theta}^{\theta+2\pi} y(t) dt \simeq u(\theta),
  \eqno(A3) 
$$
which determines the physical meaning of the function $u(\theta)$. 

We have from Eqs.~(A2): 
$$
  \left\{   \begin{array}{ll} 
  dy/d\theta = u' -a'\cos\theta + a\sin\theta - b'\sin\theta - 
  b\cos\theta, & \\ 
  d\tau/d\theta = u + \theta u' -a'\sin\theta - a\cos\theta - 
  b'(1-\cos\theta) - b\sin\theta. & 
  \end{array}    \right. 
  \eqno(A4)
$$
According to the above condition, we can neglect here terms with 
derivatives, i.e. put 
$$
  \left\{   \begin{array}{ll} 
  u' -a'\cos\theta - b'\sin\theta = 0, & \\ 
  \theta u' -a'\sin\theta - b'(1-\cos\theta) = 0, & 
  \end{array}    \right. 
  \eqno(A5)
$$
so that Eqs.~(A4) take the form: 
$$
  \left\{   \begin{array}{ll} 
  dy/d\theta = a\sin\theta - b\cos\theta, & \\ 
  d\tau/d\theta = u - a\cos\theta - b\sin\theta = y. & 
  \end{array}    \right. 
  \eqno(A6)
$$ 
Dividing the top of Eqs.~(A6) by the bottom gives: 
$$
  y\, \frac{dy}{d\tau} = a\sin\theta - b\cos\theta. 
  \eqno(A7)
$$
Moreover, Eqs.~(A6) shows that the derivatives with respect to $\tau$ 
and with respect to $\theta$ are connected by the relation 
$$
  \frac{d}{d\tau} = \frac{1}{d\tau/d\theta}\cdot \frac{d}{d\theta} = 
  \frac{1}{y}\, \frac{d}{d\theta}\,.
  \eqno(A8) 
$$ 
Applying this operator to Eq.~(A7), taking into account the first of 
Eqs.~(A2) and the condition of smallness of derivatives, we find:  
$$ 
  \begin{array}{ll}
  y\,\frac{d}{d\tau} \left( y\, \frac{dy}{d\tau} \right) = 
  \frac{d}{d\theta} (a\sin\theta - b\cos\theta) \simeq & \\ 
  a\cos\theta + b\sin\theta = u(\theta) - y. & \\ 
  \end{array}  
  \eqno(A9)
$$ 
Thus, 
$$
  y \left[ \frac{d}{d\tau} \left( y\,\frac{dy}{d\tau}\right) + 
  1 \right] = u(\theta),
  \eqno(A10)
$$ 
which coincides with Eq.~(A1) provided that 
$$
  u(\theta) = q(\tau).
  \eqno(A11)
$$
Here it suffices to restrict ourselves to the first term in the 
representation of $\tau$ from Eqs.~(A2), so that the function 
$u(\theta)$ is found from the implicit equation 
$$
  u = q(\theta \cdot u).
  \eqno(A12)
$$  

Finally, given the known function $u(\theta)$, one can find coefficients 
$a(\theta)$ and $b(\theta)$ by solving the system of linear Eqs.~(A5) 
with respect to $a'$ and $b'$, and then integrating the results. 
We will not go into further technical details.

\end{document}